\documentclass[aps,showpacs,groupedaddress,eqsecnum,notitlepage,nofootinbib,11pt]{revtex4-2}
\usepackage{graphicx,upgreek}
\usepackage{amsmath,upgreek}
\usepackage{amsfonts}
\usepackage{amssymb}
\usepackage{enumerate, color}
\usepackage{subfigure}
\usepackage{color}
\usepackage[utf8]{inputenc}
\usepackage{float}
\usepackage[colorlinks=true,pdfstartview=FitV,linkcolor=blue,citecolor=blue,urlcolor=blue,breaklinks=true]{hyperref}
\usepackage{orcidlink}
\newcommand{\be}{\begin{equation}}

\def\lb{\left[}

\def\ls{\left(}

\def\rb{\right]}

\def\rs{\right)}
\def\fr{\frac}

\def\pa{\partial}

\def\g{\gamma}

\def\ph{\phi}

\def\m{\mu}

\def\s{\sigma}

\def\o{\omega}

\def\ba{\begin{array}}
\def\ea{\end{array}}
\def\bea{\begin{eqnarray}}
\def\eea{\end{eqnarray}}

\def\la{\label}
\def\eq#1{(\ref{#1})}
\def\text#1{{\rm #1}}
\newcommand{\ee}{\end{equation}}
\newcommand{\ben}{\begin{eqnarray}}
\newcommand{\een}{\end{eqnarray}}
\newcommand{\bes}{\begin{subequations}}
\newcommand{\ees}{\end{subequations}}

\hypersetup{%
  colorlinks = true,
  linkcolor  = blue,
  citecolor = cyan,
}

\begin{document}
\title{Effective Lifshitz black holes,   hydrodynamics, and transport coefficients in fluid/gravity correspondence}
\author{D. C. Moreira$^{1}$\,\orcidlink{0000-0002-8799-3206}}
\affiliation{$^1$Centro de Ciências, Tecnologia e Saúde, Universidade Estadual da Paraíba, 58233-000, Araruna, PB, Brazil. }
\email{moreira.dancesar@gmail.com}

\author{R. da Rocha$^{2}$\,\orcidlink{0000-0003-3978-532X}}

\affiliation{$^2$Federal University of ABC, Center of Mathematics,  Santo Andr\'e, 09210-580, Brazil.}

\email{roldao.rocha@ufabc.edu.br (corresponding author)}
\begin{abstract}
Effective Lifshitz black holes with arbitrary dynamical exponent are addressed in the fluid/gravity membrane paradigm. The transport and the response coefficients in the dual Lifshitz field theory are calculated and analyzed, including the charge diffusion constant and the shear mode damping constant, along with the shear-viscosity-to-entropy density ratio. The Kubo formula is employed to obtain the electrical DC conductivity for the gauge sector corresponding to impurity through 
the holographic linear response of gauge vector fluctuations in the Lifshitz black brane geometry.
\end{abstract}

\maketitle
\section{Introduction}

AdS/CFT provides robust tools to investigate   
  field theories in the strongly coupled regime. Their hydrodynamical infrared (IR) limit corresponds to long-length scales.  
 In the AdS codimension-one bulk, weakly-coupled gravity is dual to strongly-coupled conformal field theory (CFT) on the AdS boundary.  \cite{Maldacena:1997re}. At finite temperature, the AdS bulk geometry becomes an AdS-Schwarzschild black brane with an event horizon. The AdS--Schwarzschild black brane geometry corresponds to the low-energy limit\footnote{In this limit, the  AdS$_5\times S^5$ geometry is suppressed by the effective AdS$_5$ one.} of the metric equipping a stack of $N_c$ non-extremal  Dirichlet branes, whose dual structure is an $\mathcal{N} = 4$ super-Yang--Mills theory with gauge symmetry SU($N_c$), at finite temperature that equals the one for the Hawking radiation of the gravitational background, taking into account the 
large-$N_c$ limit, for large ’t Hooft coupling $g_{\scalebox{.65}{\textsc{YM}}}\to\infty$ \cite{Witten:1998qj,Gubser:1998bc}.  
The holographic duality conjectures that the CFT at the long-scale regime on the AdS boundary must be dictated by the near-horizon limit in the AdS bulk. 
 Black holes and black branes can  encode fluid flows on their event horizon in the membrane paradigm, whose low-energy regime is a strongly-coupled
field theory.  
Einstein's equations in the AdS bulk therefore correspond to the relativistic viscous hydrodynamics regulated by Navier--Stokes equations on the AdS boundary, constituting the fluid/gravity correspondence \cite{Bhattacharyya:2008jc,Bhattacharyya:2008kq,Haack:2008cp,Ferreira-Martins:2019svk,Pinzani-Fokeeva:2014cka,daRocha:2020jdj}. 
Conformal symmetry on the gravity side corresponds to accelerated boost symmetry in the Navier–Stokes equations.
The latter are stable, strongly hyperbolic, 
 and causal, manifesting in the thorough out-of-equilibrium scenario proposed in the  BDNK  (Bemfica-Disconzi-Noronha-Kovtun) approach \cite{Bemfica:2020xym,Bemfica:2017wps,Kovtun:2019hdm,Hoult:2020eho}. The hydrodynamical fluid lives at the AdS boundary on the cutoff manifold corresponding to the ultraviolet cutoff in the dual field theory \cite{20}. It sets in the membrane paradigm within the fluid/gravity correspondence framework, wherein black holes are analog to dissipative branes with finite temperature and entropy, endowed with
electrical resistivity and
finite surface viscosity \cite{Parikh:1997ma,Price:1986yy,hub}. 
The membrane paradigm incorporates a correspondence between  the black hole geometry itself in the bulk and  the codimension-one spacetime responding like viscous fluids, emerging at the stretched horizon, which consists of a timelike manifold slightly outside the black hole event horizon, being able to thermalize, absorb, and radiate  information.
In this way, microstates of the black hole can be seen as  dynamical degrees of freedom
related to a physical membrane infinitesimally near the black hole event horizon. 
It mimics a surrogate for global event horizons in any phenomenological description of the evolution of black
holes \cite{Susskind:1993if}.

Fluid/gravity duality 
is an efficient setup to 
compute transport and response coefficients in the dual field theory having hydrodynamical description. It can be  implemented when isometries are employed to perturb the horizon of  black brane solutions of Einstein’s field equations in AdS. Among the calculation of transport and response coefficients, one of the most celebrated results of fluid/gravity duality comprises computing the shear viscosity-to-entropy density ratio ($\eta/s$).  The shear viscosity of the dual field theory can be read off the absorption cross-section of the graviton by the black brane  \cite{kss}. 
Weakly-coupled gravity has been employed as a dual scenario to explore strongly-coupled field theories, bringing robust advances for formulating correlators and the finite-temperature behavior at strong-coupled regimes  \cite{Aharony:1999ti,Romatschke:2009ng}.  The hydrodynamical limit of the gauge/gravity duality is fruitful for investigating the quark-gluon
plasma, corroborating to experimental data at the LHC and RHIC \cite{JETSCAPE:2020shq,Parkkila:2021yha,Bernhard:2019bmu,Finazzo:2016psx,Finazzo:2014cna,daRocha:2021xwq}.  Also, 1-loop quantum corrections to $\eta/s$ have been consistently addressed, complying with phenomenology \cite{Kuntz:2022kcw,Kuntz:2019omq}.   

Efforts of modeling physical systems in condensed matter have been put forward from the dual gravity side  \cite{Hartnoll:2008vx,Kovtun:2008qy}.
Relativistic strongly-coupled CFTs incorporate a dual framework that can be probed by studying gravity in asymptotically AdS spacetimes. 
The duality has recently been extended to encompass non-relativistic theories with an anisotropic scaling symmetry, nonetheless. Lifshitz spacetimes naturally emerge in quantum phase transitions occurring in condensed matter from the quantum-field theoretical point of view in the AdS/CMT setup. At the quantum critical point separating two distinct phases at the zero-temperature state, quantum fluctuations engendering the phase transition diverge, making the physical system invariant under the rescaling
$t \mapsto \lambda^z\, t$ and $x^i\mapsto \lambda \,x^i$, where $z$ is known as the dynamical exponent, for a constant $\lambda$ \cite{Kachru:2008yh,Balasubramanian:2008dm,Brito:2019ose,Bravo-Gaete:2021kgt,Ayon-Beato:2019kmz}. 
The quantum critical point splits the phase diagram into ordered and disordered phases and encodes relevant features regarding 
transport and response coefficients \cite{Pang:2009wa,Lemos:2011gy,Bravo-Gaete:2023iry}. In general these quantities are intricate to compute due to the strongly-coupled regime.
AdS/CFT can consistently address strongly-coupled systems from the gravity dual side. However, in the AdS/CFT standard formulation, top-down models are mostly  restricted to the $z=1$ case, although values $z\neq 1$ correspond to robust pieces of the duality which have been observed from the experimental point of view in condensed matter \cite{Cubrovic:2009ye,Brito:2019ose}.

Charged black brane solutions in asymptotically Lifshitz spacetimes lacked in the literature, comprising the dual gravity account of field theories at Lifshitz fixed points at finite charge density and temperature. A promising candidate was obtained in Ref. \cite{Moreira:2023zrl}, whose thermodynamic and hydrodynamic properties were thoroughly scrutinized. The solution was derived in the context of the recent extensions of Derrick's theorem for curved spacetimes \cite{alestas2019evading,carloni2019derrick,Mandal:2021qih}. One of the ways to implement it consists of  using explicitly coordinate-dependent scalar potentials, which breaks diffeomorphism invariance and yields effective stable scalar fields on static backgrounds that do not undergo backreaction \cite{bazeia2003new}. In this context, analytical scalar field solutions in the probe regime were obtained for Lifshitz and asymptotically Lifshitz spacetimes \cite{moreira2022analytical,moreira2022erratum,moreira2022scalar}, as well as for static and spherically symmetric backgrounds \cite{Morris:2021nue,moreira2023localized,Lessa:2023dbd}.     General Relativity appears as an emergent theory that manifests as the low-energy limit of some fundamental theory,  which does not necessarily obey invariance  under diffeomorphism   \cite{anber2010breaking}. The explicit  dependence of the action upon spacetime coordinates encodes degrees of freedom that are not dynamical, breaking general covariance and inducing a non-conserved  energy-momentum tensor. This fact is problematic in systems where it is necessary to consider metric backreaction since the Einstein tensor is conserved, implying a solenoidal energy-momentum tensor on shell, permitting to solve Einstein's field  equations \cite{Bluhm:2014oua}. However, in extensions of the studies carried out on the probe fields mentioned above for setups involving metric backreaction, charged black hole solutions were found with background geometry yielding asymptotically  Lifshitz spacetimes \cite{Moreira:2023zrl} and will be here explored in the fluid/gravity correspondence, emulating black brane solutions \cite{Abdalla:2009pg,daRocha:2012pt}. 

On the other hand, Lifshitz black 
brane solutions emerge as near-horizon geometries of some holographic
superconductors and AdS$_4$ charged 
black branes \cite{Gubser:2009cg,Goldstein:2009cv,Brynjolfsson:2009ct}. The Lifshitz black brane solution has also been proposed in top-down holographic models to scrutinize strange metals \cite{Hartnoll:2009ns}. 
Transport and response coefficients, as the DC conductivity and diffusion constant, can be obtained for Lifshitz black branes \cite{Taylor:2015glc}, emulating 
results from relativistic
effective field theories \cite{Kovtun:2008kx,DeFelice:2014bma}. For the non-relativistic Lifshitz black brane,  the diffusion constant and the shear viscosity-to-entropy density can be derived alternatively  from the membrane paradigm~\cite{Kovtun:2003wp,daRocha:2023waq}.

In this work,  the membrane paradigm will be employed to compute the charge diffusion constant, the shear mode damping
constant, and, subsequently, the shear-viscosity-to-entropy density ratio, as well as the electrical DC conductivity in the Lifshitz-type background geometry implemented by effective Lifshitz black hole solutions. The correlation functions can be determined by the standard procedure proposed in Refs. \cite{Son:2002sd,Policastro:2002se, Herzog:2002fn}, for arbitrary $z$ and arbitrary spacetime dimension $D$. Also, we will study the Kubo formula
driving the DC conductivity carried by impurity in Lifshitz matter. In condensed matter physics,  impurity plays an important role in approaching semiconductors. It is then essential to describe  the electrical DC conductivity also for the case of impurity. 
This work is organized as follows: in Sec. \ref{ii}, effective Lifshitz black hole solutions are briefly reported and discussed. 
Sec. \ref{iii} is devoted to presenting the fluid/gravity correspondence framework, for transport and response coefficients to be computed. The linear response against small perturbations of the  fluid membrane event horizon is implemented for the 
boundary Lifshitz dual field theory. Correlation functions yield the charge diffusion constant, the shear mode damping
constant, and the shear-viscosity-to-entropy density ratio. In Sec. \ref{iv},  the electrical DC conductivity in the Lifshitz-type background geometry, related to impurity, is implemented by fluctuations coupled to the dilaton in effective Lifshitz black hole backgrounds. Hence,  the Kubo formula is utilized, establishing 
the temperature-dependence of the electrical DC conductivity for a particular value of the dynamical exponent, corroborating with existing data in the literature. The correlation functions are derived in the context of gauge vector fluctuations, whereas impurity can be implemented as an additional gauge sector. Sec. \ref{v} contains the concluding remarks,  further discussion, and perspectives. 

\section{Effective Lifshitz black hole}
\label{ii}
In this work, we are interested in studying systems modeled by the effective action \cite{Moreira:2023zrl}
\begin{eqnarray}\label{action}
S&=&\int {\rm d}^{D} x\sqrt{-g}\left(\frac{1}{2}R-\Lambda_D-\frac{1}{2}\nabla_a \upphi\nabla^a\upphi-{\scalebox{0.99}{$V$}}(x,\upphi)-\frac{1}{2}{\scalebox{0.99}{$F_{ab}F^{ab}$}}-\frac{1}{2}\varepsilon(x){\scalebox{0.99}{$\mathcal{F}_{ab}\mathcal{F}^{ab}$}}\right),
\end{eqnarray}
for $\mathsf{g}=\text{det}(\mathsf{g}_{ab})$ denoting the metric determinant, with background coordinates $x^a$, where $a=0,1,\cdots,D-1$, $\Lambda_D=-\left(D-2\right)\left(D+3z-4\right)/2\ell^2<0$ stands for the cosmological constant, $R$ denotes the Ricci scalar,
$F_{ab}=\nabla_{[a} A_{b]}$ is the standard 
Maxwell electromagnetic field strength,  and $\mathcal{F}_{ab}=\nabla_{[a}
\mathcal{B}_{b]}$ is an additional electromagnetic field strength governing the dynamics of an auxiliary gauge vector field potential  $\mathcal{B}_a$, which interacts with a nondynamical dielectric field $\varepsilon(x)$. In addition, the scalar field $\upphi(x)$ self-interaction is governed by the coordinate-dependent scalar potential ${\scalebox{0.99}{$V$}}(x,\upphi)$. 

The {\it ansatz} for the background geometry is given by
\begin{eqnarray}
\label{backmetric}
ds^{2}=\mathsf{g}_{00}(r)dt^{2}+\mathsf{g}_{rr}(r)dr^{2}+\frac{r^2}{\ell^2}\sum_{a,b=2}^{D-1}\hat{\upsigma}_{ab}\,dx^adx^b,
\end{eqnarray}
for 
    \begin{eqnarray}
 \label{bm1} \mathsf{g}_{00}(r)&=& - \left(\frac{r}{\ell}\right)^{2z}f(r)~~~~~~\text{and}~~~~~~
\mathsf{g}_{rr}(r)=\frac{\ell^2}{r^2f(r)},
\end{eqnarray}
where  $\ell$ is a length scale, and $x^0=t$, $x^1 = r$. One assumes a  horizon metric $\hat{\upsigma}_{ab}(x^c)$, for $2\leq c\leq D-1$, endowing a transverse $(D-2)$-dimensional Einstein manifold $\hat{\Upsigma}_{\kappa}$ encoding either spherical, planar or hyperbolic topologies, for $\kappa=1$, $\kappa=0$, or $\kappa=-1$, respectively. In this way, the conserved charges, respectively associated with the Maxwell electromagnetic potential and the auxiliary gauge vector field potential $\mathcal{B}_a$, read
\begin{subequations}
\begin{eqnarray}
{\scalebox{0.99}{$Q$}}&=&-\frac{1}{4\pi}\oint_{\partial\Upsigma_\kappa} {\rm d}^{\small{D-2}}x \sqrt{|h^{(2)}|} n^a s^b {\scalebox{0.99}{$F_{ab}$}}=\frac{\upomega_{D-2}^{(\kappa)}}{4\pi\ell}q,\\[3pt]
{\scalebox{0.99}{$\mathring{Q}$}}&=&-\frac{1}{4\pi}\oint_{\partial\Upsigma_\kappa} {\rm d}^{\small{D-2}}x \sqrt{|h^{(2)}|} n^a s^b \varepsilon (x){\scalebox{0.99}{$\mathcal{F}_{ab}$}}=\frac{\upomega_{D-2}^{(\kappa)}}{4\pi\ell}\mathring{q},~~~~~~~~
\end{eqnarray}
\end{subequations}
 where $n_a$ and $s_a$ are timelike and spacelike unit normal vectors to the manifold  $\partial\Upsigma_\kappa$ defined at constant slices $(t,r)$, equipped with the induced metric $h_{ab}^{(2)}=\frac{r^2}{\ell^2}\hat{\upsigma}_{ab}$, and $\upomega_{D-2}^{(\kappa)}=\oint_{\Hat{\Upsigma}_\kappa}{\rm d}^{D-2}x\sqrt{|\Hat{\upsigma}_\kappa|}$ denotes the volume form of  $\hat{\Sigma}_{\kappa}$. 

The explicit coordinate dependence presented in the scalar potential and in the effective dielectric breaks diffeomorphism invariance and leads to a non-conserved energy-momentum tensor unless the following compatibility condition is satisfied on shell:
\begin{equation}\label{boundeq}
    \partial_a {\scalebox{0.99}{$V$}}(x, \upphi)=-\frac{1}{2}{\scalebox{0.99}{$\mathcal{F}_{bc}\mathcal{F}^{bc}$}}\partial_a \varepsilon(x).
\end{equation}
Eq. \eqref{boundeq} provides a constraint that  eliminates the conserved charge associated with  $\mathcal{B}_{a}$ from the metric solution. Therefore, the standard  Maxwell field is the only one responsible for providing electric charge to the background. The solutions for the metric coefficients, the dilaton, the electromagnetic potential, and the dielectric are respectively given by 
\begin{subequations}\label{bhsol}
\begin{eqnarray}
\label{metricsol} f(r)&=&1-2m\left(\frac{\ell}{r}\right)^{D+3z-4}+\frac{(D-3)\kappa}{(D+3z-6)r^2}+\frac{2 q^2}{\left(D-2\right)\left(D-3z\right)}\left(\frac{\ell}{r}\right)^{2(D-2)},\\[3pt]
\label{sfs} \upphi(r)&=&\varphi\pm\sqrt{(z-1)(D-2)}\log\left(\frac{r}{\ell}\right),\\[3pt]
\label{aru}A(r)&=&\Upphi-\frac{q}{D-z-2}\left(\frac{\ell}{r}\right)^{D-z-2},\\[3pt]
\label{dielfunc} \frac{1}{\varepsilon(r)}&=&\frac{\left(D+z-2\right)}{2\mathring{q}^2}\left(z-1\right)\left(\frac{r}{\ell}\right)^{2(D-2)}f(r),
    \end{eqnarray}
\end{subequations}
where $\left(\varphi,\Upphi\right)$ are integration constants\footnote{The $\Upphi$ parameter in the electromagnetic potential \eqref{aru} plays the role of the chemical potential in the near-boundary domain of the field-operator dictionary, which is going to be explored in the next sections.}.  
The solution \eqref{bhsol} is well behaved for $D\neq 3z$, $D\neq3(2-z)$, and $D\neq z+2$. Note that to achieve Lifshitz spacetime in the asymptotic regime $r\to\infty$, one must set $D+3z-4>0$. The scalar potential is given by
\begin{equation}\label{1potmodel}
{\scalebox{0.99}{$V$}}(r,\upphi)=\frac{1}{2f(r)}\left(\frac{\ell}{r}\right)^{2(D+z-2)}\left(\frac{dW}{d\upphi}\right)^{2}+U(r), ~~\text{with}~~ U(r)=-\frac{z\left(z-1\right)}{2\ell^2}f(r),
\end{equation}
and $W=W(\upphi)$ must satisfy the first-order differential equation
\begin{equation}\label{1ordereq}
\frac{d\upphi}{dr}=\pm \frac{1}{f(r)} \left(\frac{\ell}{r}\right)^{D+z-1}\frac{dW}{d\upphi}. 
\end{equation}
\textcolor{black}{Unlike previous models where general covariance is respected and ensures the compatibility of the field equations with the contracted Bianchi identity, the above solution is derived from the effective action \eqref{action}, which explicitly breaks general covariance, emerging in setups where we can - in addition to choosing a suitable structure for the scalar potential - simplify the constraint \eqref{boundeq} and deal with it within the field equations of the model. In this way, the solution \eqref{bhsol}  has also an effective character, implying the naming of effective black holes.} By assuming that the conditions for the existence of an event horizon $(r_{\textsc{h}})$ are satisfied, one can write the mass parameter as
\begin{eqnarray}
    m(r_{\textsc{h}})=\frac{1}{2}\left(\frac{r_{\textsc{h}}}{\ell}\right)^{D+3z-4}\left[1+\frac{(D-3)\kappa}{(D+3z-6)r_{\textsc{h}}^2}+\frac{2 q^2}{\left(D-2\right)\left(D-3z\right)}\left(\frac{\ell}{r_{\textsc{h}}}\right)^{2(D-2)}\right].
\end{eqnarray}
In this system, one can find scenarios presenting either a unique event horizon for $z>D/3$ or two horizons (an event horizon and a Cauchy horizon) for $1\leq z\leq D/3$ \cite{Moreira:2023zrl}.

\section{Charged planar Lifshitz black brane and the membrane paradigm}
\label{iii}
Several hydrodynamical features of the effective Lifshitz black hole solutions can be probed in the membrane paradigm setup, wherein the stretched horizon
can be transliterated into a dissipative viscous fluid with electrical conductivity and shear viscosity. 
The analog hydrodynamic description holds for the effective Lifshitz black hole solution, whose horizon lacks diffeomorphism invariance. This analogy will also be used in the next section for establishing the mapping between the bulk fields in the action (\ref{action}) and the dual field theory currents in the long-wavelength regime.
The approach to black hole gravitational perturbations indicates that fluctuations of the stretched horizon have to encode the complete collection of
hydrodynamical modes, {which  correspond to a shear wave and a sound wave propagating in a conformal plasma in thermal equilibrium}. Methods in fluid/gravity correspondence must replicate wave modes describing the sound propagation, in addition to non-propagating modes including shear and diffusion.
Hydrodynamics, as an effective theory, designates  the dynamics ruling any thermal system whose time and length scales are large, contrasted to microscopic scales. The degrees of freedom entering this theory, in the simplest
cases, are the densities of conserved charges.
Hydrodynamics proposes observing a shear diffusion mode whose diffusion constant is proportional to the shear viscosity. 
We will compute, for charged  Lifshitz black branes, for an  arbitrary dynamical exponent $z$, the charge diffusion constant, the shear mode damping
constant and, subsequently, the shear-viscosity-to-entropy density ratio \cite{Iqbal:2008by,Pang:2009wa}. 

One can now take into account the charged  Lifshitz black brane metric \eqref{backmetric}, with coefficients (\ref{bm1}, \ref{metricsol}), for the planar case corresponding to $\hat{\upsigma}_{ij} = \delta_{ij}$ in the metric \eqref{backmetric}.  {\color{black}{Here we aim to emulate the fluid/gravity approach to charged black brane solutions in asymptotically Lifshitz spacetimes. Therefore, the planar solutions $\kappa=0$ are the only option, as long as the  low-energy, long-wavelength limit corresponding to viscous  hydrodynamics can be implemented in the field theory which is dual to generalized gravity described by the action \eqref{action}, in fluid/gravity correspondence. In fact, the near-horizon limit of a stack of $D_3$-branes in string theory is equivalent to the metric of a class of black holes with planar horizon, in the low-energy regime, which can be seen as an appropriate limit of the standard Lifshitz black hole geometry \cite{Korovin:2013bua}. Besides, the coordinates labelling the gauge theory spatial coordinates   on $\mathbb{R}^3$  correspond to the planar horizon (regarding to $\kappa=0$) of the black brane, which implements the gravity dual to the gauge theory. 
One can add perturbations to black holes and take the low-energy, long-wavelength limit for the hydrodynamic analysis of large-$N_c$ plasmas. 
For black holes with planar horizons, one can split the perturbation as plane waves, and the low-energy, long-wavelength limit for viscous  hydrodynamics can be implemented. However, for a black hole with a compact horizon, corresponding to  $\kappa=1$, the perturbation wavelength   automatically becomes the black hole scale.  For black holes with  spherical horizons,  perturbations are decomposed into spherical harmonics, which present a discrete spectrum and the hydrodynamical limit cannot be implemented \cite{Natsuume:2014sfa}. 
 Therefore, the hydrodynamical limit of AdS/CFT, encompassing fluid/gravity correspondence,
can only be naturally implemented for black holes with planar horizon.
The charged effective Lifshitz black hole with planar horizon, corresponding to $\kappa=0$, is the natural option for a holographic dual fluid/gravity correspondence in asymptotically Lifshitz spacetimes. 
An additional reason for considering planar horizons corresponding to $\kappa=0$ is to consider appropriate limits of the standard Lifshitz black hole solutions to recover the 
AdS-Schwarzschild black brane, which has planar horizon itself. 
The historical origin of the importance of black holes with planar horizons comes from 
Ref. \cite{Witten:1998zw}, which considered a maximally supersymmetric Yang--Mills theory on the $p + 1$ dimensional  world-volume of $N_c$ $D_p$-branes 
\cite{Witten:1998zw,Aharony:1999ti}. For the particular case where $p=3$, the solution of the $D_3$-brane corresponds to the $\mathcal{N} = 4$ super-Yang--Mills theory at zero temperature describing fluctuations of the branes \cite{Tong:2005nf,Horowitz:1998ha}. 
}}

Equations of motion governing gauge vector fields, with suitable  boundary conditions, 
yields the dispersion relation for the diffusive mode, $\omega=\omega(k)= -i\mathfrak{D} k^2$, taking the limit $k\to0$, encoding the charge diffusion constant for black brane backgrounds. 
In the membrane paradigm, a conserved current $j^a$ can be defined in terms of the field strength 
$F_{ab}$ appearing in the action \eqref{action} \cite{Kovtun:2003wp,
Iqbal:2008by,Parikh:1997ma}.
The stretched horizon is a  spacelike surface situated at $r=r_0$ such that $
  r_{\textsc{h}} < r_0$ and $r_0-r_{\textsc{h}} \ll r_{\textsc{h}}.$
The outward normal to that surface is a unit spacelike vector $n_a=(0,\mathsf{g}^{1/2}_{rr},0, \cdots, 0)$.
The current associated with the stretched horizon in the  membrane paradigm reads 
$
  j^a = \lim_{r\to r_0} n_b F^{ab}$, whereas the  parallelizability of the current to the horizon, $n_a j^a=0$ and, additionally, current conservation  
$\partial_a j^a=0$, reside on the fact that the field  strength $F^{ab}$ is antisymmetric. Ref. \cite{Kovtun:2003wp} showed Fick's law $j^a=-\mathfrak{D} \partial_a j^0$, as the current conservation equation $\partial_a j^a = 0$ implies that the charge density $j^0$ satisfies the diffusion equation 
$\left(\partial_t -  \mathfrak{D} \, \mathsf{g}^{ab}\partial_a \partial_b\right)j^0$ = 0.

Taking into account the metric (\ref{backmetric}) with coefficients (\ref{bm1}, \ref{metricsol}) the charge diffusion constant reads  
\begin{eqnarray}
\label{2eq6}
\mathfrak{D}&=&-\frac{\ell^2\sqrt{-\mathsf{g}(r_{\textsc{h}})}}{\sqrt{-\mathsf{g}_{00}(r_{\textsc{h}})\mathsf{g}_{rr}(r_{\textsc{h}})}}
\int^{\infty}_{r_{\textsc{h}}}dr\frac{\mathsf{g}_{00}(r)\mathsf{g}_{rr}(r)}{\sqrt{-\mathsf{g}(r)}}\nonumber\\
 &=&\frac{r_{\textsc{h}}^{2 D+2}}{({D-z-2}){\ell}^{D+z-2}},
\end{eqnarray}
where the inequality $z<D-2$ is assumed to ensure the convergence of the integral. 

On the other hand, 
the dispersion relation for the shear mode reads  $\omega=\omega(k)= -i\mathcal{D} k^2$, arising from the lowest pole remaining in the long-wavelength, low-frequency hydrodynamic limit, due to the black hole graviton absorption upon gravitational perturbations. 
One interprets the diffusion as a consequence of viscosities of the
dual gauge theory plasmas also in the limit $k\to0$, where the shear mode damping constant is given by 
$\mathcal{D}=\eta/(\upepsilon+P)$.
The shear viscosity comes upon the response of the fluid,  dual to the effective Lifshitz black hole solution,  undergoing thermal internal forces. By coupling gravity to the hydrodynamical fluid and measuring the response of the energy-momentum tensor under gravitational perturbations, the Kubo formula established the shear viscosity as a transport coefficient associated with the  retarded Green function \cite{Natsuume:2014sfa,Ferreira-Martins:2019wym}. 
For gravitational perturbations of the effective Lifshitz black hole solution, the shear mode damping
constant is given by~\cite{Kovtun:2003wp,
Iqbal:2008by}
\begin{equation}
\label{2eq10}
\mathcal{D}=-\frac{\sqrt{-\mathsf{g}(r_{\textsc{h}})}}{\sqrt{-\mathsf{g}_{00}(r_{\textsc{h}})\mathsf{g}_{rr}(r_{\textsc{h}})}}\int^{\infty}_{r_{\textsc{h}}}
dr\frac{\ell^2 \mathsf{g}_{00}(r)\mathsf{g}_{rr}(r)}{r^2\sqrt{-\mathsf{g}(r)}}=-\frac{r_{\textsc{h}}}{2 D-z-2} \left(\frac{r_{\textsc{h}}}{\ell}\right)^{-D+z-1}.
\end{equation}

Ref. \cite{Moreira:2023zrl} showed that the temperature of the effective Lifshitz black hole solution \eqref{backmetric}, with coefficients (\ref{bm1}, \ref{metricsol}), is given by 
\begin{eqnarray}\label{bht}
  T_{\textsc{h}}&=&\frac{r_{\textsc{h}}^z}{4\pi \ell^{z+1}}\left(D+3z-4+\frac{(D-3)\kappa}{r_{\textsc{h}}^2}-\frac{2q^2}{D-2}\frac{\ell^{2(D-2)}}{r_{\textsc{h}}^{2(D-2)}}\right).
\end{eqnarray}
One can therefore derive the expression for $\eta/s$ by observing the  thermodynamic expression $\epsilon+P=T_{\textsc{h}}s$, where $s$ stands for the entropy per unit volume of the fluid membrane,  yielding 
\ben
\frac{\eta}{s}&=&T_{\textsc{h}}\frac{\sqrt{-\mathsf{g}(r_{\textsc{h}})}}{\sqrt{-\mathsf{g}_{00}(r_{\textsc{h}})\mathsf{g}_{rr}(r_{\textsc{h}})}}\int^{\infty}_{r_{\textsc{h}}}
dr\frac{-\mathsf{g}_{00}(r)\mathsf{g}_{rr}(r)}{\mathsf{g}_{xx}(r)\sqrt{-\mathsf{g}(r)}}\nonumber\\
&=&\frac{1}{4\pi(2D-z-2)}\left(\frac{r_{\textsc{h}}}{\ell}\right)^{-D+2z}\left(D+3z-4+\frac{(D-3)\kappa}{r_{\textsc{h}}^2}-\frac{2q^2\ell^{2(D-2)}}{(D-2)r_{\textsc{h}}^{2(D-2)}}\right),\label{kssl}
    \een
 where the integral is convergent if $z<D-2$. Eq. (\ref{kssl}) emulates the Kovtun-Son-Starinets (KSS) result for charged planar Lifshitz black branes \cite{kss,Son:2002sd}.

\section{Asymptotic expansion and electrical DC conductivity for the impurity sector}
\label{iv}

Einstein-Maxwell-dilaton theory, with a
Liouville-type potential is associated with a relativistic theory with non-conformal symmetry  
\cite{Kulkarni:2012re}. In this case, the electrical DC conductivity is temperature dependent if vector fluctuations couple to the dilaton. The Lifshitz geometry can be recovered if a Liouville 
potential is absent \cite{Kachru:2008yh,Taylor:2015glc} and consists of the weakly-coupled dual to a strongly-coupled Lifshitz field theory. The bulk vector field endowing the Lifshitz geometry is dual to matter in the Lifshitz field theory and the limit  $z\to 2$ recovers the non-relativistic nature of the dual theory. The electrical DC conductivity was investigated, also in the context of superconductivity, in  nonrelativistic 
Lifshitz media in the absence of dilaton couplings \cite{Ge:2010aa,Mann:2009yx}. 
However, the DC conductivity in the Lifshitz medium was also computed in the context of  non-vanishing dilaton couplings.
We here aim to implement vector fluctuations in the effective Lifshitz black hole geometry,  
to report impurity in the Lifshitz medium. 
The fluctuation of the background gauge field can be associated with 
the Lifshitz matter, whose DC conductivity is different from the one related to impurity. 
Macroscopic features of non-relativistic Lifshitz media can probed by methods involving the linear response theory provided by  fluctuations. 

The electric properties of a non-relativistic Lifshitz theory with two different charge carriers can be studied when one  perturbatively implements impurities into the action (\ref{action}). In the charged planar Lifshitz black brane scenario, in the decoupling limit $g_{\scalebox{.65}{\textsc{YM}}}^2 \gg 1$ of weakly-coupled gravity, for the particular case $z=1$, the scalar and gauge sectors become decoupled from gravity and do not contribute to the AdS$_D$ curvature. It is reasonable to work with an equivalent decoupling limit for any value of $z$ for the charged planar Lifshitz black brane. Therefore, to analyze impurities, all analytical and numerical methods can be used in the fixed background of the charged planar Lifshitz black brane described by the  metric \eqref{backmetric}, with coefficients (\ref{bm1}, \ref{metricsol}). The action (\ref{action}) can be perturbatively added by the action describing impurity as \cite{Pang:2009wa,Park:2013goa}
\be
S_{\textsc{imp}} = -  \int {\rm d}^D x \sqrt{-\mathsf{g}}  \  
 \fr{1 }{4} e^{\g \upphi}  \mathsf{H}_{ab} \mathsf{H}^{ab},
\ee
where   $\mathsf{H}_{ab} =\pa_{[a} B_{b]}$, 
where  $B_{a}$ is an Abelian gauge field potential, with  dilaton coupling different from $A_{a}$ and $\mathcal{B}_a$.
The coupling between the vector fluctuations and the dilaton is regulated by the parameter $\g$. 
On the gravity side of the holographic duality,
described by the effective Lifshitz solution,  the two aforesaid gauge vector field potentials $A_{a}$ and $\mathcal{B}_a$
correspond to different matter fields portraying Lifshitz matter, whereas $B_a$ describes impurity, which is assumed to not interact with the background gauge field in quadratic order. Therefore it does not mix  
with metric fluctuations  \cite{Pang:2009wa,Lee:2010qs}. It is worth mentioning that the two-gauge current model coupled with gravity and a dilaton in AdS/CMT has been used also in describing 
graphene near the charge-neutrality point, composing the 
Dirac fluid of a strongly-interacting plasma in hydrodynamics. An additional gauge field was proposed to study the thermal conductivity and the electrical DC conductivity in graphene \cite{Seo:2016vks}. Ref. \cite{Rogatko:2017tae} also employed a two-current  holographic model permitting one of them 
to carry the dark
matter sector in graphene. Besides, the holographic gauge/gravity duality leading generalized to Navier--Stokes equations and horizons with soft-hair excitations  was  employed to study impurity in graphene \cite{Ferreira-Martins:2021cga}.

Denoting  by $\mathsf{b}_{a}$ the fluctuations associated with $B_{a}$, implementing effects of  impurity in the dual Lifshitz field theory,   one can consider  the part of the action responsible for fluctuations, as
\bea  
S_{\textsc{b}} &=& -  \int {\rm d}^D x \frac14\sqrt{-\mathsf{g}}  \  e^{\g \upphi}   \mathsf{h}_{ab} \mathsf{h}^{ab},
\la{act:fluctb}
\eea  
where  $\mathsf{h}_{ab} = \pa_{[a} \mathsf{b}_{b]}$. The electrical DC conductivity carried by impurity can be then obtained with the aid of the Kubo formula. 
From the action \eq{act:fluctb}, the transverse mode, $\mathsf{b}_c$, where $c$ denotes transverse coordinates,   satisfies the linearized equation of motion \cite{Pang:2009wa,Park:2013goa} 
\be\label{eq:impurityeq}
\pa^{\m} \lb \sqrt{-\mathsf{g}} e^{\g \ph}\, \mathsf{g}_{ac}\, \mathsf{g}_{i d}\,
\pa_{[c} \mathsf{b}_{d]} \rb = 0.
\ee
The Fourier transform 
$
\mathsf{b}_c (\omega,r) = \int \fr{d \o}{2 \pi} e^{ i \o t} \ \mathsf{b}_c (t,r)$ 
at the horizon  has two solutions, to wit
\be 		\la{sol:nearhsol}
\mathsf{b}_c  (\omega,r) = a_1 f^{ \pm \upnu}(r) ,
\ee
taking into account the metric coefficient  \eqref{metricsol}, with $\upnu = i \fr{\o}{4} r_{\textsc{h}}^{-D-z+4}$, where $a_1$ is a suitable constant.  The sign in the exponent of Eq. (\ref{sol:nearhsol}) corresponds to the ingoing/outgoing 
boundary conditions at the horizon. Picking an ingoing wave,  
the solution of \eq{eq:impurityeq} in the hydrodynamic limit $\o \ll T$ can be expressed as 
\be		\la{sol:psolb}
\mathsf{b}_c (r) =  f^{ - \upnu}(r) \lb H_0 (r) + \o H_1 (r)    \rb  + {\cal O} (\o^2) .
\ee
where $H_0 (r)$, $H_1(r)$ must be regular at the horizon. Eq. (\ref{sol:psolb}) must be led to Eq. \eq{sol:nearhsol} at the horizon. Therefore,  $H_0 (r)$ must attain the value of a constant $a_1$ at the horizon $r_{\textsc{h}}$, whereas $H_1 (r_{\textsc{h}})$ and higher-order terms must vanish therein.
These conditions yield
\bea
\!\!\!\!\!\!\!\!\!\!\!\!\!\!\!\!H_0 (r) &=& a_1 \!+\!\int_{r_{\textsc{h}}}^r\frac{ {\rm r}^{-D-3 z+3}}{2 q^2 {\rm r}^4 \left(\frac{\ell}{{\rm r}}\right)^{2 D}\!+\left[D^2\!+\!6 z\!-\!D (3 z\!+\!2)\right]\left[\ell^4\!-\!r_{\textsc{h}}^{3 z} \left(\frac{\ell^4 q^2}{r_{\textsc{h}}^4 (4\!-\!3 z)}+1\right)
   \left(\frac{\ell}{{\rm r}}\right)^{D+3 z} {\rm r}^4\right] }d{\rm r}, \la{res:udicon} \\
\!\!\!\!\!\!\!\!\!\!\!\!\!\!\!\!\!\!H_1 (r) &\!=\!& a_2  \!+\!i\frac{a_1}{ r_{\textsc{h}}^{D \!+\!  z \!-\! 4}}\log\left[\frac{q^2}{r_{\textsc{h}}^{4-3z} (4\!-\!3 z)}
\left(\frac{\ell}{r}\right)^{D+3 z-4}\!\!\!\!\!\!\!+\!\frac{2
   q^2 \left(\frac{\ell}{r}\right)^{2 D-4}}{(D-2) (D-3 z)}\right]\left[1-\left(\frac{r}{r_{\textsc{h}}}\right)^{D +  z - 4}\right]\nonumber\\&&
-a_3\int_{r_{\textsc{h}}}^r \frac{e^{3 {\rm r}} {\rm r}^{-2 \gamma -z+1}}{{\rm r}^4 r_{\textsc{h}}^{3 z-4} \left[q^2+r_{\textsc{h}}^4 (4-3 z)\right]
   \left(\frac{\ell}{{\rm r}}\right)^{D+3 z}-q^2 r_{\textsc{h}}^4 \left(\frac{\ell}{{\rm r}}\right)^{2 D}-\ell^4 (4-3 z)}\,d{\rm r},
\eea
with
\bea
a_2 &=& \frac{i a_1}{8}
\left[1-\left(\frac{r}{r_{\textsc{h}}}\right)^{D +  z - 4}\right]\frac{B\left[\beta(r);\frac{1}{2}\left(\gamma+D-2z\right), \left(\frac{r_{\textsc{h}}}{r}\right)^{D+z-2}\right]}{B\left[\frac{1}{2}(\gamma+D-2z), \left(\frac{r_{\textsc{h}}}{r}\right)^{D+z-2}\right]},
  \eea
denoting by  ${B}[x;\,\alpha,\beta]=\int _{0}^{x}t^{\alpha-1}\,(1-t)^{\beta-1}\,dt$  the incomplete beta function and by ${B}[\alpha, \beta]=\frac {\Gamma (\alpha)\,\Gamma (\beta)}{\Gamma (\alpha+\beta)}$ the standard beta function, where  
\bea
\beta(r) = \frac{r^5 r_{\textsc{h}}^{3 z-4} }{D+3
   z-5}\left[q^2+r_{\textsc{h}}^4 (4-3 z)\right]\left(\frac{\ell}{r}\right)^{D+3 z}+\frac{q^2 r^5}{5-2 D} \left(\frac{\ell}{r}\right)^{2 D}+\ell^4 r (4-3 z),
\eea 
 and 
\bea
a_3 &=& - i a_1 \left(\frac{r_{\textsc{h}}}{\ell}\right)^{-(D +  z - 4) \g}.
\eea
The constant term $a_1$ can be determined as long as one imposes an ancillary 
boundary  condition. At the asymptotic boundary, the vector fluctuation driving impurity  reads
\be			\la{res:asymtsoldu1}
\mathsf{b}_c (r) = \mathsf{b}_1  + \mathsf{b}_2 \ r^{(D +  z - 4)\g-2} ,
\ee
where $\mathsf{b}_1$ is a source term, whereas $\mathsf{b}_2$ stands for the VEV of the dual operator. In the range $\g < 1$, the asymptotic behavior of the vector fluctuation is driven by $\mathsf{b}_1$ and the following Dirichlet boundary condition can be assumed \cite{Park:2013goa}: 
\be			\la{123}   \,
\mathsf{b}_1  =  \lim_{r_c \to \infty}  \mathsf{b}_c  (r_c) ,
\ee
where $r_c$ implies an appropriate UV cutoff of the dual theory.
When one compares the asymptotic expansion of the perturbative solution  \eqref{sol:psolb} with Eq.  \eqref{123}, it yields 
\be			\la{bval:imp}
a_1 = \fr{8 i r_{\textsc{h}}^{D +  z - 4}}{\tau(r_{\textsc{h}})}\,\mathsf{b}_1 ,
\ee
where 
\begin{eqnarray}
\tau(r_{\textsc{h}}) &=& 8 i r_{\textsc{h}}^{D +  z - 4} +\o  \left[{\rm Ein} \ls \frac{1}{2}(\gamma+D-2z+1)\rs - {\rm Ein}  \ls \frac{1}{2}(\gamma+D-2z+1) \rs\right. \nonumber\\&& \qquad\quad\qquad\quad\qquad\quad\left.+ \tan^{-1} \frac{2\sqrt{3}}{3} -\frac{\pi}3 + 2 \pi \log \ls \frac{\pi}{2}(\gamma+D-2z+1) \rs\right],  
\end{eqnarray}
and ${\displaystyle \operatorname {Ein} (x)=\Gamma (0,x)+\upgamma +\log x}$ is the entire exponential integral,   $\Gamma (0,x)={ \int _{x}^{\infty }t^{-1}\,e^{-t}\,dt}$ is the incomplete gamma function and $\upgamma=\lim _{n\to \infty }\left(-\log n+\sum _{k=1}^{n}{\frac {1}{k}}\right)$ stands for the Euler--Mascheroni constant.

The boundary action corresponding to the on-shell action of \eq{act:fluctb} is given by
\begin{eqnarray}
\!\!\!S_{\textsc{boundary}} &=& -  \int\limits_{r \to r_c} {\rm d}^{D-1} x \ \sqrt{-\mathsf{g}}  \ e^{\g \upphi} \mathsf{g}^{rr} \mathsf{g}^{cc}
\mathsf{b}_{c}  \mathsf{b}_c^\prime\nonumber\\  &\approxeq&   \int {\rm d}^{D-1} x \ r_{\textsc{h}}^{- (D +  z - 4)\g } \
r_c^{D-1 - (D +  z - 4)\g}   \ \mathsf{b}_{1}\mathsf{b}_c^\prime \label{bdy}.
\end{eqnarray}
Hence, normalizable contributions to $S_{\textsc{boundary}}$ in Eq. (\ref{bdy}) have $\mathsf{b}_c^\prime  \sim r_c^{1-D + (D +  z - 4)\g}$
as the leading term, in the limit $r_c \to \infty$.
Since the asymptotic expansion takes leading-order terms from Eq. \eqref{sol:psolb} as
\be
{\mathsf{b}_c^\prime} = - \fr{i a_1 \o}{r_{\textsc{h}}^{ (D +  z - 4) \g}}  {r_c^{1-D + (D +  z - 4)\g}} + {\cal O} \ls \fr{1}{r^{D+z-1}} \rs ,
\ee
the current-current retarded Green function \cite{Policastro:2002se,Pang:2009wa,Park:2013goa} yields\footnote{Hereon we reinstate the $\frac{1}{16\pi G_D}$ terms, omitted heretofore due to the use of natural units throughout the text. We denote by $G_D$ the $D$-dimensional gravitational coupling constant.}
\be\label{grii}
G_R^{cc}(\omega,k) =  -i\int {\rm d}^Dx\, e^{-ik\cdot x}
\theta(t)\left\langle\left[J^c(x), J^c(0)\right]\right\rangle = \fr{i \o}{16 \pi G_D} \fr{1}{r_{\textsc{h}}^{(D +  z - 4)\g}} + {\cal O} (\o^2) ,
\ee
where \eq{bval:imp} was used and $\theta(t)$ is the Heaviside step function. The electrical DC conductivity, from the Kubo formula, therefore reads
\be			\la{res:kubob}
\s_{\textsc{dc}}= -\lim_{\substack{{\omega}\rightarrow 0\\
 k\to0}} \frac{ \Im G_R^{cc} ({\omega},{k})}{\omega} = \frac{1}{16 \pi G_D} \fr{1}{r_{\textsc{h}}^{(D +  z - 4)\g}},
\ee
carrying the response of the electric current under perturbations in the electric field, which is generated by the gauge vector potential.

Eq. (\ref{bht}) regards the temperature of the effective Lifshitz black brane. For $D=4$, it generalizes the well-known expression
$T_{\textsc{h}}=\frac{z+2}{4\pi}r_{\textsc{h}}^z$ valid for standard Lifshitz black branes \cite{Pang:2009wa},   
whose expression can be alternatively obtained by expanding the metric near the horizon, lacking any conical singularity.
For the case $z=2+\epsilon$ and $D=4$, one can invert Eq. (\ref{bht}), considering a planar horizon ($\kappa=0$) in Eq. (\ref{bht}), and writing the non-negative solution for the horizon radius in terms of the temperature, as
\begin{eqnarray}
\lim_{\epsilon\to0}r_{\textsc{h}}=\frac{2\sqrt{3}\ell}{3}\sqrt{\sqrt{2 \pi ^2 T_{\textsc{h}}^2 \ell ^2-3 q^2}+2 \pi  T_{\textsc{h}} \ell}.
\end{eqnarray}
Therefore we obtain, 
\be			\la{res:kubob1}
\s_{\textsc{dc}}  \propto{T_{\textsc{h}}^{-\gamma}},
\ee
  recovering the electrical DC conductivity dependence with the temperature and the $\gamma$ parameter controlling the coupling between the dilaton and the additional gauge vector, regulating impurity. 
It shows consistency with  results in Ref. \cite{Park:2013goa}. 
Hence, our results seem to corroborate with the ones in gauge/gravity, as the correlator in the dual Lifshitz field theory is regulated by bulk field fluctuations near the asymptotic boundary, for the sector controlling impurity. 

\section{Ending comments and concluding remarks} 
\label{v}

The membrane paradigm of fluid/gravity was employed to investigate the hydrodynamics underlying effective Lifshitz black hole solutions. 
In particular, the charge diffusion constant, the shear mode damping constant, and  the shear-viscosity-to-entropy density ratio were computed and discussed for effective Lifshitz black hole solutions, generalizing scenarios in the literature involving standard Lifshitz black branes. The Kubo formula,  relating  transport coefficients to appropriate components of the Green function, was handled  to compute the electrical DC conductivity for the gauge sector corresponding to impurity. This method consistently posed the holographic linear response of gauge vector  fluctuations describing impurity in the effective Lifshitz black brane geometry.  The analog of the KSS result was also addressed in the effective Lifshitz black hole scenario.  The charge diffusion constant and the electrical DC conductivity are obtained from the correlators using Kubo formul\ae\,,  corroborating with the membrane paradigm for arbitrary dimension and dynamical exponent. The usual dependence of the electrical DC conductivity with the black hole temperature was recovered for particular cases of the spacetime dimension and dynamical exponent. 

As perspectives, one can implement second-order hydrodynamics for the effective Lifshitz black hole solution, taking into account vorticity, to compute the relaxation time, the gravitational susceptibility, and the shear relaxation \cite{Finazzo:2014cna}. Also, one can explore anisotropic Weyl anomalies in Lifshitz gauge/gravity scenarios. In standard AdS/CFT, the Weyl anomaly 
comes from breaking the conformal invariance of classical fields under Weyl symmetries after the quantization procedure and naturally underlies black hole solutions. 
Comparing the holographic Weyl anomaly to the
trace anomaly of the energy-momentum tensor of the dual field theory yields a measure of the backreaction of black branes onto the bulk geometry of Lifshitz spacetimes, emulating well-known results regarding the AdS bulk geometry \cite{Meert:2021khi,Meert:2020sqv}. Hence Weyl and trace anomalies can further determine the accuracy of describing effective Lifshitz black branes in the membrane paradigm of AdS/CFT. One can emulate and adapt this construction for the anisotropic Weyl anomaly and for addressing asymptotic expansions for holographic renormalization \cite{Griffin:2012qx}.

\subsection*{Acknowledgments}
R.d.R.~thanks to The S\~ao Paulo Research Foundation -- FAPESP
(Grants No. 2021/01089-1 and No.~2022/01734-7), and to the National Council for Scientific and Technological Development -- CNPq  (Grants No. 303742/2023-2 and No. 401567/2023-0), for partial financial support.

\bibliographystyle{iopart-num}

  \bibliography{biblio} 
\end{document}